\documentstyle[a4,11pt]{article}

\newcommand{\nit}{\noindent}

\newcommand{\np}{\newpage}
\newcommand{\dsp}{\displaystyle}
\newcommand{\vs}[1]{\vspace{#1 ex}}
\newcommand{\hs}[1]{\hspace{#1 em}}
\newcommand{\bfr}{\begin{flushright}}
\newcommand{\efr}{\end{flushright}}
\newcommand{\bc}{\begin{center}}
\newcommand{\ec}{\end{center}}
\newcommand{\ben}{\begin{enumerate}}
\newcommand{\een}{\end{enumerate}}

\newcommand{\be}{\begin{equation}}
\newcommand{\ee}{\end{equation}}
\newcommand{\ba}{\begin{array}}
\newcommand{\ea}{\end{array}}
\newcommand{\ct}{\cite}
\newcommand{\bit}{\bibitem}
\newcommand{\dd}[2]{\frac{\partial{#1}}{\partial{#2}}}

\newcommand{\ve}{\varepsilon}

\newcommand{\lb}{\lambda}

\newcommand{\fg}{\phi}
\newcommand{\vf}{\varphi}
\newcommand{\og}{\omega}

\newcommand{\Del}{\Delta}

\newcommand{\Og}{\Omega}

\newcommand{\lh}{\left(}
\newcommand{\rh}{\right)}

\begin{document}

\pagestyle{empty} 
\begin{flushright}
NIKHEF/02-010
\end{flushright} 
\vs{3}
\begin{center} 
{\Large{\bf  Cosmological Higgs fields}} \\
\vs{7} 

{\large J.W.\ van Holten$^*$ }\\
\vs{2} 

{\large NIKHEF, Amsterdam NL}\\ 
\vs{4}

September 16, 2002 
\vs{15} 

{\small{\bf Abstract}}
\end{center} 

\nit
{\footnotesize{We present a time-dependent solution to the coupled Einstein-Higgs 
equations for general Higgs-type potentials in the context of flat FRW cosmological 
models. Possible implications are discussed. }} 
\vfill
\footnoterule 
\nit
{\footnotesize{$^*${\tt{e-mail: v.holten@nikhef.nl}} \\
Work supported by the program FP52 of the Foundation for Research of Matter (FOM) 
}}
\np

~\hfill
\np

\pagestyle{plain}
\pagenumbering{arabic} 

\nit 
Scalar fields play a fundamental role in the standard model of particle physics, as well 
as its possible extensions. In particular, scalar fields generate spontaneous symmetry 
breaking and provide masses to gauge bosons and chiral fermions by the Brout-Englert 
mechanism \ct{be} using a Higgs-type potential \ct{higgs}. As observed by Linde 
\ct{linde} and Veltman \ct{veltman2}, the scalar-field energy condensed in the vacuum 
contributes to an effective cosmological constant, with a typical value many orders of 
magnitude larger than observed. At the same time, the cosmological effects of scalar 
fields have been proposed as a mechanism to drive the evolution of the universe in 
various scenarios \ct{guth,linde2,wetterich,peebles,steinhardt1}; for reviews, see 
e.g.\  \ct{lyth,bahcall,sahni,straumann}. The nature of the proposed cosmological 
scalars is unknown. It is the purpose of this paper to investigate the minimal 
coupling of Higgs scalars to gravity and discuss the solutions from the point of 
view of cosmology as well as particle physics. 

We consider a flat FRW-type universe ($k = 0$) with scale factor $a(t)$, and a 
set of minimally coupled scalar fields $\fg_i$. In the limit in which matter and 
radiation can be neglected, but maintaining the homogeneity and isotropy of the 
universe, the dynamics of this system is governed by the equations 
\be 
\ba{l} 
\dsp{ \frac{1}{2}\, \sum_i \dot{\fg}_i^2\, + V[\fg_i]  = \frac{3H^2}{8\pi G}, }\\ 
 \\
\dsp{ \ddot{\fg}_i + 3 H \dot{\fg}_i + V_{,i} = 0,} 
\ea 
\label{1}
\ee 
where $H = \dot{a}/a$ is the Hubble parameter.  Constant scalar fields
minimizing the potential solve these equations for constant Hubble parameter:
\be 
H = H_0 = \sqrt{\frac{8\pi G V_0}{3}}.
\label{2}
\ee 
Such solutions of eqs.(\ref{1}) are relevant in the context of the standard model 
of particle physics, and its supersymmetric and/or gauge-unified extensions, in 
which space-time is taken to be static and globally Lorentz invariant. In fact, in flat 
space-time $H_0 = V_0 = 0$, which requires a careful tuning of the parameters in the 
theory. From observations we know, that global Lorentz invariance is actually violated 
by the expansion of the universe\footnote{And locally of course by such gravitational 
fields as those of the earth and the sun.}. However, the expansion is very slow and the 
measured present value of the Hubble parameter $H_0 \approx 70$ km/sec/Mpc 
corresponds to an extremely low energy density $V_0$ of about $5$ GeV/m$^3$. 
This is of the order of $10^{-123}$ in units of Planck energy per Planck volume, or 
$10^{-45}$ in terms of a typical QCD energy density. Therefore the approximation 
$H_0 = 0$ is excellent on scales relevant to particle physics experiments, and no 
violations of Lorentz invariance have been observed there. In particular the tuning of 
parameters in the Higgs potential of the standard model is not affected by the observed 
expansion of the universe. 

In contrast, the expansion of the universe and the associated non-zero energy densities 
are relevant in a cosmological context. To model the cosmological behaviour of Higgs-type 
scalars we consider a single field component inducing spontaneous symmetry 
breaking; this field is denoted by $\vf = \fg_1$. We take all other scalar fields $\fg_i$ 
to vanish in the vacuum state: $\fg_i = 0$ $(i \neq 1)$. The Higgs potential $V(\vf) = 
V[\fg_1 = \vf, \fg_i = 0]$ then reduces to a quartic polynomial of the form 
\be 
V(\vf) = \ve + \frac{m^2}{2}\, \vf^2 + \frac{\lb}{4}\, \vf^4.
\label{3}
\ee 
For $m^2 = - \mu^2 < 0$ the minimum of the potential is in the regime of spontaneous 
symmetry breaking,  characterized by  
\be 
\vf_0 ^2 = \frac{\mu^2}{\lb}, \hs{2} V_0 = \ve - \frac{\mu^4}{4\lb}. 
\label{4}
\ee 
This is consistent with eqs.(\ref{1}) only for $\ve \geq \mu^4/4\lb$. In particular 
a static Lorentz-invariant universe requires the parameters of the potential to be 
related by $\ve = \mu^4/4\lb$. 

With effectively a single minimally coupled scalar field $\vf$, the cosmological Einstein-Higgs 
equations (\ref{1}) can be recast into the form 
\be 
\frac{1}{2}\, \dot{\vf}^2 = - \frac{\dot{H}}{8\pi G}, \hs{2} 
V(\vf) = \frac{3H^2 + \dot{H}}{8\pi G}. 
\label{5}
\ee 
Assuming the existence of a well-defined solution in some finite period of time $\Del t$, 
we can equivalently take the Hubble parameter to be a function of $\vf(t)$: $H(t) = 
H[\vf(t)]$. Then $\dot{H} = H^{\prime} \dot{\vf}$, where the prime denotes a derivative 
w.r.t.\ $\vf$, and it follows that either $\vf$ is constant, or we have non-static fields 
($\dot{\vf} \neq 0$) satisfying the equations  
\be
4\pi G \dot{\vf} = - H^{\prime}, \hs{2} 8\pi G V(\vf) = 3H^2 - \frac{H^{\prime\, 2}}{4\pi G}.
\label{6}
\ee 
Comparing with the expression (\ref{3}), we see that the expressions match if and only if 
$H[\vf]$ is of the form 
\be 
H = h + 2 \pi G \og \vf^2,
\label{7}
\ee 
where $h$ and $\og$ are constant parameters with the dimension of inverse time.  
The matching again requires a relation between the three parameters $(\ve, \mu^2, \lb)$, 
implicitly given by  
\be 
\ve = \frac{3h^2}{8\pi G}, \hs{2} 
m^2  = -\mu^2  = \og^2 \lh \frac{3h}{\og} -1 \rh, \hs{2} \lb = 6\pi G \og^2. 
\label{8}
\ee 
Using expression (\ref{7}) for $H$, the first equation (\ref{6}) then determines the 
time dependence of the scalar field:
\be 
\dot{\vf}= - \og \vf \hs{1} \Rightarrow \hs{1} \vf(t) = \vf(0)\, e^{- \og t}. 
\label{9}
\ee 
The explicit form of the constraint (\ref{8}) is
\be 
\ve = \frac{\mu^4}{4\lb}\, \lh 1  - \frac{\lb}{6\pi G \mu^2} \rh^2,
\label{8.1}
\ee
which is the direct analogue of the relation $V_0 = 0$ above for the existence of 
a constant solution in a static and Lorentz-invariant universe. This constraint is 
quite interesting; for instance, we observe that for $\mu^2 > 0$ and
\be
0 < \lb < \frac{3 \mu^2}{2 M_{Pl}^2}, \hs{3} M^2_{Pl} = \frac{1}{8 \pi G},
\label{8.2}
\ee 
the value of the vacuum energy $\ve$ for the dynamical scalar field (\ref{9}) is lower 
than that for any static field solution (\ref{4}). In contrast, in the regime without 
static symmetry breaking ($m^2  = - \mu^2 > 0$), the dynamical solution always 
requires a value $\ve > 0$, unlike Minkowski space-time with vanishing 
Higgs field and $\ve = 0$. Thus in this class of models, for given $\mu^2$ and 
$\lb$ it is the value of $\ve$ which determines which solution is actually realized.

As concerns the scalar field $\vf(t)$, for $\og > 0$ it vanishes asymptotically, even 
in the case of a potential with non-trivial minima ($m^2 < 0$). In contrast, for 
$\og < 0$ the scalar field evolves away from this symmetry point. For the evolution 
of the Hubble parameter and the scale factor the scalar field solution (\ref{9}) 
results in 
\be 
H(t) = h + 2 \pi G \og \vf^2(0)\, e^{-2\og t}, \hs{2} 
a(t) = a(0) e^{h t + \pi G \vf^2(0) \lh 1 - e^{-2\og t} \rh}. 
\label{1.0}
\ee
For the particular case $h = \ve = 0$ these solutions are well-known and have been 
used e.g.\ in the context of chaotic inflation models \ct{linde2}. 

For $\og > 0$ and at times $t > 1/\og$ the solution (\ref{1.0}) describes standard 
exponential expansion, with constant Hubble parameter $h$; if this situation is to 
describe the observed universe, $h$ must be small. For early times the scale factor 
grows faster, with initial Hubble parameter $H_1 \approx h + 2\pi G \og \vf^2(0)$. 
In contrast, for $\og < 0$, the Hubble parameter itself decreases exponentially fast 
until it vanishes, when $a(t)$ reaches its maximum; subsequently $H(t)$ becomes 
negative and the universe starts to contract in a super-exponential way. 

For the further analysis it is convenient to introduce dimensionless fields $\chi$ 
and decay parameter $x$:
\be 
\chi^2 = \frac{8\pi G}{3}\, \vf^2, \hs{2} x = \frac{\og}{h}. 
\label{1.1}
\ee 
Then for $h \neq 0$ we can define a dimensionless potential 
\be 
\Og_V = \frac{8\pi G V}{3 h^2}\, = 1 - \frac{1}{2}\, x(x - 3) \chi^2 
 + \frac{9x^2}{16}\, \chi^4. 
\label{1.2}
\ee 
which expresses the potential energy density in terms of the asymptotic critical energy 
density. For $\og > 0$ this potential has a stable minimum at $\chi^2 = 0$ in the range 
$0 \leq x \leq 3$, and non-trivial minima for $\chi^2 \neq 0$ in the domain $x > 3$; 
$\og < 0$  implies $x< 0$, with non-trivial minima only. 

Another quantity of interest is the parameter $N = 3\chi^2(0)/8$; the dynamical solutions 
for the field and scale factor can then be written as 
\be 
\chi(t) = \chi(0) e^{-xh t}, \hs{2} a(t) = a_0 e^{ht + N (1 - e^{-2xht})}.
\label{1.3}
\ee 
Furthermore, the Hubble parameter and the potential vary in time as 
\be 
H(t) = h \lh 1 + 2x N e^{-2xh t} \rh, \hs{1} 
\Og_V = 1 - \frac{4}{3}\, (x-3) xNe^{-2xht} + 4x^2 N^2 e^{-4xht}.  
\label{1.4}
\ee 
If $h$ is close to the asymptotic value of $H(t)$, then 
\be 
H(t = h^{-1}) \approx h \hs{1} \Rightarrow \hs{1} e^{2x} > 2xN,
\label{1.4.1}
\ee  
which can always be satisfied for large enough $x$. 
The quantity $N$ has a simple interpretation: it represents the extra number of 
$e$-folds by which the universe inflates between the initial time and the moment 
at which the expansion becomes dominated by the asymptotic Hubble constant $h$. 
Obviously during this period the expansion of the universe is much faster than 
a pure de Sitter expansion with constant Hubble parameter $h$. For $N$ to be 
larger than one, the initial scalar field must take values of the order of the 
Planck scale: 
\be 
\vf^2(0) = \frac{N}{\pi G}\, = N \lh 0.7 \times 10^{19}\, \mbox{GeV} \rh^2. 
\label{1.5}
\ee 
In the early universe such large-amplitude scalar fields may have existed, 
e.g.\ in the symmetry breaking sector of a unified gauge theory; the classical 
treatment is still considered reliable, as long as the scalar potential remains 
well below the Planck scale \ct{linde2}. For example, scalar neutrino fields 
of Planck-scale magnitude have been proposed in the context of supersymmetric 
leptogenesis models \ct{yanagida}. Note that time dependent scalar fields would help 
to establish non-equilibrium and non time-reversal invariant conditions in the epoch 
of lepto- and baryogenesis. 

The coupling of Higgs scalars to vector fields in this scenario induces a time-dependent 
mass for the vector bosons. With a minimal coupling between scalar and vector field, 
the vector field equation in the Lorenz gauge in a gravitational and scalar background 
(\ref{9}), (\ref{1.0}) becomes 
\be 
\lh \frac{1}{a^2}\, \dd{}{\eta} a^2 \dd{}{\eta}  - \Del + 2 g^2 |\vf|^2 \rh A_{\mu} = 0,
\label{1.6}
\ee 
where $\eta$ is the conformal time defined by $d\eta = dt/a(t)$. The mass gap is then 
given by 
\be
M^2(t) = 2g^2 |\vf|^2(t) = \frac{2Ng^2}{\pi G}\, e^{-2xh t} \hs{1} \Rightarrow \hs{1}
 \frac{M^2(t = h^{-1})}{M^2(0)}\, = e^{-2x}. 
\label{1.7}
\ee 
If the initial value of the vector boson mass is of the order of the Planck mass  
$M^2(0) = M^2_{Pl}$, then $ 16 N g^2 = 1$. Such an initial value is consistent with a late 
value $M_{GUT} \approx 10^{16 }$ GeV for $x \approx 6$. The inequality (\ref{1.4.1}) 
then only imposes a very mild restriction on $N$: $N < 1.4 \times 10^{4}$.

It is to be observed that if $h = H_0$, the present Hubble parameter, and with $x$-values 
of the order $10$ or less, the energy density represented by the potential (\ref{1.2}) 
indeed remains small even if $N$ is large: at most a few orders of magnitude more than 
the present critical density. Hence it can not dominate the energy density in the early 
universe, and our starting eqs.(\ref{1}) presumably  can hold only at relatively late 
times. In contrast, if we take the initial scalar energy density to be the Planck density,
then for $h = H_0$ and $N > 1$ we obtain $xN \approx 10^{\,61}$, or equivalently 
\be 
\og \approx  \frac{10}{N \tau_{Planck}}, \hs{2} \tau_{Planck} = 0.53 \times 10^{-43}\, 
 \mbox{sec}.  
\label{1.8}
\ee 
Such a scalar field disappears within $N$ Planck times; however, the
inflation by $N$ $e$-folds would also happen within this same period. 

In the Higgs models defined by (\ref{1.2})  there is no compelling reason why $h$ should 
be taken to represent the presently observed value of the Hubble parameter; actually it 
represents the asymptotic value of $H(t)$ in the regime where the eqs.(\ref{1}) are 
relevant, which may be only in the very early universe. In such a scenario $h$ is determined 
by the critical density at the beginning of the radiation dominated era, rather than by 
the present critical density; the asymptotic critical density would then naturally be of 
the order of the GUT scale. 

A special class of potentials arises for  $x = 3$. In this case the Higgs mass 
term vanishes and the number of parameters is reduced. This model was studied 
as a flat-space quantum field theory by Coleman and Weinberg \ct{coleman}. The 
potential $\Og_V$  in (\ref{1.4}) then reaches its asymptotic value in a time 
scale $t = h^{-1}$ if 
\be 
\Og_V (t = h^{-1}) = {\cal O}(1) \hs{1} \Leftrightarrow \hs{1} 
6N e^{-6} = 1,
\label{1.9}
\ee 
which gives $N \approx 70$, as in standard models of inflation \ct{guth}. For $h$ in 
the range of the GUT-scale, the epoch of inflation is sufficiently short: depending on the 
exact unification scale in the range $\tau = h^{-1} \approx 10^{\,6 - 8}\, \tau_{Planck}$. 
For the evolution of the vector boson masses we now obtain $M(t = h^{-1}) \approx M(0)/20$,
which puts the initial unified gauge boson masses at the string scale, rather than the 
Planck scale.  
 
In conclusion, the combined FRW-Einstein-Higgs equations (\ref{1}) allow constant
as well as time-dependent solutions for all values of $(\mu^2,\lb)$; which solution
is realized depends on the value of $\ve$. The dynamical solutions which satisfy 
(\ref{8.1}) may find application in models of the early universe to bridge the gap 
between the GUT- and Planck scales, to assist in baryogenesis, or to generate inflation. 
As soon as matter and radiation contribute substantially to the energy density 
eqs.\ (\ref{1}) are modified. The scalar fields can then become thermalized and 
finite temperature corrections to the potential have to be included as well (see 
e.g.\ \ct{lyth} and references therein). In a non-equilibrium situation more work 
is needed to study time-dependent Higgs fields in the presence of macroscopic 
densities of matter and radiation \ct{jwvh}.

\end{document}